\documentclass[12pt]{article}
\usepackage{amssymb,amsmath,amsfonts,amssymb}
\usepackage{cite}
\newlength{\dinwidth}
\newlength{\dinmargin}
\newtheorem{theorem}{Theorem}[section]
\newtheorem{prop}[theorem]{Proposition}
\newtheorem{lemma}[theorem]{Lemma}

\newenvironment{proof}{\medskip \noindent 
            {\bf Proof.}}{ \hfill $\square$ \medskip}

\newcommand{\ie}{{\it i.e.\ }}

\newcommand{\bp}{{\mbox{\boldmath $p$}}}
\newcommand{\sbp}{{\mbox{\footnotesize \boldmath $p$}}}

\newcommand{\bx}{{\mbox{\boldmath $x$}}}
\newcommand{\sbx}{{\mbox{\footnotesize \boldmath $x$}}}

\def\idty{{1}}
 
\def\net{{\Rs}}
\def\hatnet{{\widehat{\Rs}}}

\def\wrnet{{\{\Rs(W)\}_{W \in\Ws}}}

\def\As{{\cal A}}
\def\Bs{{\cal B}}
\def\Cs{{\cal C}}

\def\Fs{{\cal F}}

\def\Hs{{\cal H}}
\def\Is{{\cal I}}

\def\Ms{{\cal M}}
\def\Ns{{\cal N}}
\def\Os{{\cal O}}
\def\Ps{{\cal P}}

\def\Rs{{\cal R}}
\def\Ss{{\cal S}}

\def\Ws{{\cal W}}

\def\Pid{{\Ps_+ ^{\uparrow}}}

\def\RR{{\mathbb R}}
\def\CC{{\mathbb C}}
\def\NN{{\mathbb N}}
\def\IN{{\mathbb N}}

\def\bx{{\mbox{\boldmath{$x$}}}}

\def\beq{\begin{equation}}
\def\eeq{\end{equation}}
\begin{document}

\title{String-- and Brane--Localized Causal Fields \\ 
in a Strongly Nonlocal Model} 

\author{{\Large Detlev Buchholz\,$^a$ \ and \  
Stephen J.\ Summers\,$^b$ }\\[5mm]
${}^a$ Institut f\"ur Theoretische Physik, 
Universit\"at G\"ottingen, \\ 37077 G\"ottingen, Germany  \\[2mm]
${}^b$ Department of Mathematics, 
University of Florida, \\ Gainesville FL 32611, USA}

\date{}

\maketitle 

{  \abstract \noindent 
We  study a weakly local, but nonlocal model in
spacetime dimension $d \geq 2$ and prove that it is maximally
nonlocal in a certain specific quantitative sense. Nevertheless, 
depending on the number of dimensions $d$, it 
has string--localized or brane--localized 
operators which commute at spatial distances.
In two spacetime dimensions, the model  
even comprises a covariant and local subnet of
operators localized in bounded subsets of Minkowski space which has
a nontrivial scattering matrix. The model thus exemplifies the algebraic
construction of local operators from algebras associated with
nonlocal fields.}

\section{Introduction}

\setcounter{equation}{0}

 Until recently, locality has been viewed as a basic feature 
of quantum field theory. However, nonlocal theories arising naturally
in the context of quantum gravity and string theory \cite{ElWo},
in curved and noncommutative space--times \cite{AdS,DFR,DoNe}, 
and in recent approaches to the construction of local theories 
\cite{Sch1,Sch2,AdS,LoRe,BuLe,Le,Le2,Le3},
throw a new light on this matter. In this paper we examine a prototype 
of a nonlocal theory in $d \geq 2$ spacetime dimensions from
the point of view of these and other recent developments. The model is based
on a weakly local but nonlocal quantum field, which satisfies all remaining
Wightman axioms and the superstability conditions studied in
\cite{AdS}.

     In particular, we are interested in
investigating just how nonlocal this model is and to which extent
there are remnants of locality which have physical significance.  We
shall show that the model is maximally nonlocal in a specific,
quantitative sense. Nonetheless, it contains ``string--localized''
or -- depending on the number of spacetime 
dimensions -- ``brane--localized'' 
operators which commute at spatial distances.  In two spacetime
dimensions, it contains a local, covariant net for which a full
scattering theory can be defined, yielding a scattering matrix not
equal to the identity. This illustrates in another concrete model a recently
emerged approach \cite{Sch1,Sch2,AdS,LoRe,BuLe,Le,Le2,Le3} to establish the
existence of local theories by starting from 
algebras generated by nonlocal operators.  
The essential advantage of this novel approach is that the 
intricate explicit construction 
of local operators can be avoided by using algebraic techniques. 

     After introducing the model in the next section, we 
prove in Section 3 that it is maximally nonlocal. In Section 4 we
compute its
modular structure and establish some immediate consequences.  We
then investigate the independence properties of spacelike separated
algebras in Section 5. In Section 6 we show that,
depending on the 
number of spacetime dimensions, the nonlocal 
wedge algebras generated by the field contain point--, string-- 
respectively brane--localized operators which satisfy the
condition of locality. In the final section, we discuss the
significance of these findings from a number of vantage points and
make some further comments.

\section{The model}  

\setcounter{equation}{0}

     The model we are studying in this paper has been known at least
since R.\ Jost's classic monograph on axiomatic quantum field theory
\cite{Jost}. It describes a scalar massive Fermion and was used to 
establish that weak local commutativity of field operators is a strictly 
weaker condition than local commutativity in the context of Wightman's 
axioms. 

We shall consider this model in Minkowski space--time of any dimension 
$d \ge 2$. It is defined  on the antisymmetric
Fock space $\cal H$ over the one--particle space  ${\cal H}_1$ of
a scalar particle of mass $m > 0$, \ie the direct sum
of the antisymmetrized $n$-fold tensor products 
${\cal H}_n = {\cal H}_1 \wedge \cdots \wedge {\cal H}_1$, $n \in \NN$,  
and the space ${\cal H}_0$ consisting of multiples of the
vacuum state $\Omega$. We identify 
${\cal H}_1$ with the $L^2$--space of momentum--space wave functions 
equipped with the standard Poincar\'e invariant scalar product and
consider the linear mapping of the space of test functions
$\Ss(\RR^d)$ into  ${\cal H}_1$ given by restricting the 
Fourier transforms $\widetilde{f}$ of $f \in \Ss(\RR^d)$ to the
positive mass shell $M$ 
$$  |f \rangle  \doteq \widetilde{f} \upharpoonright 
\{  p \in \RR^d : p_0 = \sqrt{\bp^2 + m^2} \} \, .
$$
The scalar product of $ |f \rangle,  |g \rangle \in {\cal H}_1$ is given by
$$
\langle f | g \rangle = 2 \pi \! \int \! dp \, \theta(p_0) \delta(p^2 - m^2)
\overline{\widetilde{f}(p)} \widetilde{g}(p) \, .
$$
The natural action of the Poincar\'e transformations 
$\lambda \in \Pid$ on the test functions, 
$(\lambda f) (x) \doteq f(\lambda^{-1}x), \, x \in \RR^d$,
induces a continuous unitary representation $U$ of 
$\Pid$ on $\cal H$ which leaves $\Omega$
invariant and acts on the generating vectors according to
$$
U(\lambda) \, | f_1 \rangle \wedge \cdots \wedge | f_n \rangle =
| \lambda f_{1} \rangle \wedge \cdots \wedge | \lambda f_{n} \rangle 
\, .
$$
The representation $U$ satisfies the relativistic spectrum 
condition (positivity of the energy in all Lorentz frames).

On the antisymmetric 
Fock space $\Hs$ one can introduce in a standard manner a 
field $\phi$ satisfying canonical anticommutation relations. Concretely,
$\phi$ is the operator-valued distribution
$$
\phi : \Ss(\RR^d) \longrightarrow \Bs(\Hs) \, ,
$$
which acts on the generating vectors according to 
\begin{equation} \label{fieldaction}
\begin{split}
& \phi(f) \,  | f_1 \rangle \wedge \cdots \wedge | f_n \rangle \\ 
& = |f\rangle \wedge  | f_1 \rangle \wedge \cdots \wedge | f_n \rangle +
\sum_{k =1}^n (-1)^{k+1} \, \langle \overline{f} | f_k \rangle
\,  | f_1 \rangle \wedge \cdots 
\overset{k}{\vee} \cdots \wedge | f_n \rangle \, ,
\end{split}
\end{equation}
where $\overset{k}{\vee}$ denotes omission of $|f_k \rangle$.
It follows from this definition 
that $\phi$ satisfies the Klein--Gordon equation
$$  
\phi\big( (\square + m^2)f \big) = 0 \, ,
$$
the anticommutation relations
$$
\big\{ \phi (f), \phi (g) \big\} \doteq \phi (f)  \phi (g) +  \phi (g)  \phi (f) =
\big( \langle \overline{g} | f \rangle +
\langle \overline{f} | g \rangle \big) \cdot \idty \, ,
$$
and the hermiticity condition 
$$   \phi(f)^* = \phi(\overline{f}) \, .
$$
Moreover, the unitary representation $U$ of the Poincar\'e group acts
covariantly on the field,
$$
U(\lambda) \phi(f) U(\lambda)^{-1} = \phi(\lambda f) \, .
$$
Thus $\phi$ is a scalar Fermi field which, in accordance with the 
Spin--Statistics--Theorem, is nonlocal.

As is well known, one can decompose the field $\phi$ into creation
and annihilation parts, $\phi(f) = \phi_+(f) + \phi_-(f)$, whose
respective actions on the generating vectors are given by the first
and second term on the right hand side of equation (\ref{fieldaction}).
So one obtains further anticommutation relations 
$$
\big\{ \phi_+ (f), \phi(g) \big\}  =
\langle \overline{g} | f \rangle \cdot \idty \, , \quad 
\big\{ \phi_- (f), \phi(g) \big\}  =
\langle \overline{f} | g \rangle \cdot \idty \, .
$$

For our purposes here, we must introduce some further objects. Setting
$$V = (-1)^{N(N-1)/2} \, , \quad Z = (-1)^N \, ,$$
where $N$ is the particle number operator acting on the 
subspaces $\Hs_n \subset \Hs$ by multiplication with $n$, $n \in \NN_0$, 
we define an additional field 
$\widehat{\phi} : \Ss(\RR^d) \rightarrow \Bs(\Hs)$. It is given by
$$\widehat{\phi}(f) \doteq V \phi(f) V^{-1} 
= (\phi_+(f) - \phi_-(f)) \, Z \, , $$
where the latter equality follows from the fact that  
$\phi_\pm(f)$ changes the particle number by $\pm 1$ and $(-1)^{N^2} = Z$. 
Since $V$ commutes with all Poincar\'e transformations
$U(\lambda)$, the field $\widehat{\phi}$ transforms 
in the same covariant manner under the adjoint action of these transformations
as the original field $\phi$.

Although the fields $\phi$, $\widehat{\phi}$ are nonlocal, they
are relatively local, as is seen by the following computation.
Noticing that  $Z$ and $\phi$ anticommute, 
$Z \phi(g) + \phi(g) Z = 0$, one obtains for the commutator
$$
[ \widehat{\phi}(f), \phi(g) ] = 
- \{ \phi_+(f) - \phi_-(f), \phi(g) \} \, Z 
= (\langle \overline{f} | g \rangle - \langle \overline{g} | f \rangle ) \, 
Z \, .
$$
But the expression
$$
\langle \overline{f} | g \rangle - \langle \overline{g} | f \rangle 
= 
 2 \pi \! \int \! dp \,\, \varepsilon(p_0) \delta(p^2 - m^2)
\widetilde{f}(-p) \widetilde{g}(p) 
$$
is the smeared commutator distribution affiliated with the
Klein--Gordon equation, cf.\ \cite{Jost}, which vanishes if the 
supports of the test functions $f$ and $g$ are spacelike separated. 

     We consider in the following von Neumann algebras 
which are generated by the fields smeared with test functions
having support in given subregions of Minkowski space. 
First, for any nonempty open subset $\Os \subset \RR^d$, we define
$\Rs(\Os)$ to be the von Neumann algebra generated by the set
of operators 
$$
\{ \phi(f) : f \in \Ss(\RR^d), \ \textnormal{supp} \, f \subset \Os \} \, .
$$
With this definition, the map $\Rs : \Os \mapsto \Rs(\Os)$ is
clearly inclusion preserving and thus defines a net on $ \RR^d$. 
Due to the covariance properties of the field $\phi$, this net transforms
covariantly under Poincar\'e transformations, \ie
$$
U(\lambda) \Rs (\Os) U(\lambda)^{-1} = \Rs (\lambda \Os) \, ,
$$
but it is not local, \ie it is not true that the elements of the algebras 
$\Rs(\Os_1)$ and $\Rs(\Os_2)$ commute whenever the regions 
$\Os_1$ and $\Os_2$ are spacelike separated. 

The second net 
$\widehat{\Rs} : \Os \mapsto \widehat{\Rs}(\Os)$
is generated in the same way by the field $\widehat{\phi}$.
Thus, $\widehat{\Rs} (\Os) = V \Rs (\Os) V^{-1}$ for
every $\Os \subset \RR^d$, so this net is Poincar\'e covariant
and nonlocal as well. Yet, because of the relative locality
of the fields $\phi$ and $\widehat{\phi}$ established above,
the elements of the algebras $\Rs (\Os_1)$ and $\widehat{\Rs}(\Os_2)$
do commute whenever $\Os_1$ and $\Os_2$ are spacelike
separated. Denoting the spacelike complement of $\Os$ by
$\Os'$ and the commutant of $\Rs(\Os)$ by $\Rs(\Os)'$, we therefore 
have the following algebraic starting point.  

\begin{lemma} \label{2.1}
The two nets $\net$, $\hatnet$ are $U(\Pid)$-covariant, nonlocal nets
which are relatively local in the sense that one has  
$$\Rs(\Os) \subset \widehat{\Rs}(\Os ')' = V \Rs(\Os ')' V \, , $$
for any open $\Os \subset \RR^d$.
\end{lemma}

     In this paper a distinguished class of spacetime regions called 
wedges plays a special role. Choosing proper coordinates, let 
$W_R \doteq \{ x  \in \RR^d \mid x_1 > 
\vert x_0 \vert \}$ denote the ``right wedge''
and $W_L \doteq \{ x  \in \RR^d \mid - x_1 > 
\vert x_0 \vert \} = W_R^\prime$ the corresponding ``left wedge''.
For $d > 2$, the set
$\Ws$ of wedges is given by $\Ws = \{ \lambda W_R \mid \lambda \in \Pid \}$.
In two dimensions, the set $\Ws$
of wedges has two disconnected components, one consisting of
the translates of $W_R$ and the 
other of the translates of $W_L$.

\section{Degree of nonlocality of the model} \label{nonlocal}

\setcounter{equation}{0}

     In order to gauge the nonlocality of the model under
investigation, we employ a quantitative measure of commensurability
of observables recently introduced in \cite{BSloc}. 
Let $\As$, $\Bs$ be two von Neumann algebras acting on some Hilbert
space $\Hs$, and let $\As \bigvee \Bs$ be the von Neumann algebra 
generated by them. A measure of the strength of correlations between 
$\As$ and $\Bs$ in a normal state $\omega$ on  $\As \bigvee \Bs$
is given by 
\begin{equation*}
C_\omega(\As,\Bs) \doteq
\sup_{E,F} |\omega(E \wedge F) - \omega(E) \omega(F)| \, ,
\end{equation*}
where the supremum extends over all pairs of projections $E \in \As$,
$F \in \Bs$, and $E \wedge F$ is the maximal projection contained in both
$E$ and $F$. {}From this one can proceed to a measure 
of the incommensurability of the two algebras, setting 
\begin{equation*}
C(\As,\Bs) \doteq \inf_\omega \; C_\omega(\As,\Bs) \, ,
\end{equation*}
the infimum being taken over all normal states on $\As \bigvee \Bs$.

The values of $C (\As,\Bs)$ can be shown
to lie in the interval $[0,1]$, where
the extreme value $0$ implies the  
commutativity of $\As$ and $\Bs$, provided 
the algebra $\As \bigvee \Bs$ is simple,
cf.\ \cite{BSloc}. The value 1 indicates that 
the two algebras are maximally incommensurable; it is 
realized, for example, in quantum mechanics by the algebras generated 
by the position and momentum operator, respectively. 
Thus, if for spacelike separated regions $\Os_1,\Os_2$
one has $C(\Rs(\Os_1),\Rs(\Os_2)) = 1$, it is meaningful to 
say that the underlying net $\Rs$ is maximally nonlocal. We shall compute 
this invariant and show that the net in the present model
is indeed maximally nonlocal. 
As a matter of fact, this result holds also for the 
subnet $\Rs_e$ generated by the even polynomials in the field $\phi$.
 
Let $f_1,f_2$ be real test functions which are orthonormal in the sense
$$
\langle f_1 \mid f_1 \rangle = \langle f_2 \mid f_2 \rangle = 1 \, , \,\, 
\mbox{Re} \, \langle f_1 \mid f_2 \rangle = 0 \, ,
$$
so that $\phi(f_1),\phi(f_2)$ are self-adjoint involutions and
$\phi(f_1) \phi(f_2) = - \phi(f_2) \phi(f_1)$. For such test functions, 
the operator $i \phi(f_1) \phi(f_2)$ is self-adjoint and unitary; hence 
it is an involution. It follows that 
$P_\pm = \frac{1}{2}(\idty \pm i \phi(f_1) \phi(f_2))$ are orthogonal 
projections in $\Hs$ and $P_+ + P_- = \idty$. Similarly, let 
$g_1,g_2$ be another pair of such test functions and set 
$Q_\pm = \frac{1}{2}(\idty \pm i \phi(g_1) \phi(g_2))$.
We assume that the supports of  $f_1,f_2$ are
spacelike separated from those of $g_1,g_2$.

\begin{lemma} \label{3.1}
Let  $f_1, f_2$ and $g_1,g_2$ be pairs of
test functions as described above such that 
$ \langle f_1 + i f_2 | g_1 - i g_2 \rangle \neq 0$.
Then one has $P_+ \wedge Q_+ = 0$.
\end{lemma}

\begin{proof} Setting     
$S = \frac{1}{2} \, \phi(f_1 - if_2), \, T = \frac{1}{2} \, \phi(g_1 - ig_2)$,
it is straightforward to verify the equalities
\begin{equation*}
\begin{split}
& SS^* = P_+ \, , \  S^* S = P_- \, , \
T T^* = Q_+ \, , \ T^* T = Q_- \, , \\
& ST + TS = \mbox{\small $\frac{1}{2}$} \, \langle f_1 + i f_2 | g_1 - i g_2 
\rangle \cdot 1 \, .
\end{split}
\end{equation*}
In the latter equality the fact was used again that 
$\langle \overline{f} | g \rangle = \langle \overline{g} | f \rangle$, if $f$ 
and $g$ have spacelike separated supports.
Since $S = P_+ S = S P_-$ and $P_+ + P_- = 1$, one has
$S \, P_+ = S \, (1-P_- ) = 0$ and, 
similarly, $T \, Q_+ = 0$. Thus if $\Phi \in P_+ \Hs \cap Q_+ \Hs$,
it follows that $S \Phi = T \Phi = 0$ and consequently
$$
 \mbox{\small $\frac{1}{2}$} \, \langle f_1 + i f_2 | g_1 - i g_2 
\rangle \, \Phi = (ST + TS) \, \Phi = 0 \, .
$$
Hence, under the given conditions on the test functions,  
$ P_+ \Hs \cap Q_+ \Hs = \{ 0 \}$, so that
$P_+ \wedge Q_+ = 0$.
\end{proof}

Next, we construct sequences of functions with the properties
specified above, which will be used to exhibit certain 
specific central sequences of projections. 

\begin{lemma} \label{3.2}
There exists a sequence $\{ f_{1, n}, f_{2, n} \}_{n \in \NN}$
of orthonormal pairs of real test functions, 
such that $\mbox{supp}\, f_{j ,n} \subset
\{x \in \RR^d : |x| < 1/n \}$, $j = 1,2$,  and
$\lim_{n \rightarrow \infty} \, \langle f_{1, n} | f_{2, n} \rangle = -i$.
Moreover, $\lim_{n \rightarrow \infty} \, 
\langle f_{j, n} | g \rangle = 0$,
$j = 1,2$, for any test function $g$. 
\end{lemma} 

\begin{proof}
Recall that the entire analytic ``sine integral'' $\mbox{Si}$ is given by
$$ z \mapsto \mbox{Si}(z) = \frac{2}{\pi} \, 
\int_0^z \! dw \, \frac{\sin w}{w} \, , $$
where the normalization is chosen in such a way that
$\lim_{z \rightarrow \pm \infty}  \mbox{Si}(z) = \pm 1$; its Fourier transform
is a tempered distribution with support in the interval $[-1,1]$. 
Let $x \mapsto h(x)$ be any real test function on $\RR^d$ which is 
symmetric about the origin and has support
in the ball $\{x \in \RR^d : |x| < 1/2 \}$, and let 
$\widetilde{h}(0) = 0$. 
Consider corresponding 
sequences of test functions $h_{1,n}, h_{2,n}$, $n \in \NN$, given in 
momentum space by
\begin{equation*}
\begin{split}
& \widetilde{h}_{1,n}(p) \doteq (i/n^{d-2}) \,
 \mbox{Si}(p_0/2n) \, \widetilde{h}(p/n^2) \, , \\
& \widetilde {h}_{2,n}(p) \doteq (1/n^{d-2})  \,
 \widetilde{h}(p/n^2) \, .
\end{split}
\end{equation*}
Upon multiplication with suitable
normalization factors which will be determined below, 
these functions have all properties stated in the lemma. To verify 
this, begin by noting that the test functions $h_{1,n}, h_{2,n}$ 
are real. Moreover, in view of the 
support properties of $h$ and those of the Fourier transform of 
$\mbox{Si}$, the supports of $h_{1,n}$ and $h_{2,n}$ are contained in the ball 
$\{x \in \RR^d : |x| < 1/n \}$.  After a change of variables, one
obtains for the scalar products
\begin{equation} \label{scalarproducts}
\begin{split}
& \langle h_{1,n} | h_{1,n} \rangle =
2 \pi \! \int \! dp \, \theta(p_0) \delta(p^2 - m^2/n^4) \,
 |\mbox{Si}(n p_0/2)|^2 \, | \widetilde{ h}(p) |^2 \, , \\
& \langle h_{2,n} | h_{2,n} \rangle =
2 \pi \! \int \! dp \, \theta(p_0) \delta(p^2 - m^2/n^4) \,
 | \widetilde{ h }(p) |^2 \, , \\
& \langle h_{1,n} | h_{2,n} \rangle =
- 2 \pi i \! \int \! dp \, \theta(p_0) \delta(p^2 - m^2/n^4) \,
\mbox{Si}(n p_0/2) \, | \widetilde{ h }(p) |^2 \, .
\end{split}
\end{equation}
The latter relation implies that $\mbox{Re} \, \langle h_{1,n} |
h_{2,n} \rangle = 0$, $n \in \NN$. Moreover, it follows from the above 
relations and the properties of $\mbox{Si}$ that 
\begin{equation*}
\lim_{n \rightarrow \infty} \,  \langle h_{1,n} | h_{1,n} \rangle =
\lim_{n \rightarrow \infty} \,  \langle h_{2,n} | h_{2,n} \rangle =
i \lim_{n \rightarrow \infty} \,  \langle h_{1,n} | h_{2,n} \rangle =
2 \pi \! \int \! dp \, \theta(p_0) \delta(p^2) \,
 | \widetilde{ h }(p) |^2 \, , 
\end{equation*}
where one recalls that $\widetilde{h}(0) = 0$, so the integral
exists also in $d = 2$ dimensions. Choosing a function  $h$
such that this integral is different from $0$,
one concludes that the sequence of pairs
$\{ f_{1,n} \doteq \langle h_{1,n} | h_{1,n} \rangle^{-1/2} \, h_{1,n},
f_{2,n} \doteq \langle h_{2,n} | h_{2,n} \rangle^{-1/2} \, h_{2,n}
\}_{n \in \NN}$ has all properties stated in the first parts of 
the lemma. The last assertion of the lemma is an immediate consequence of the 
relations 
\begin{equation*}
\begin{split}
& \langle h_{1,n} | g \rangle =
- (2 \pi i / n^{d-2} ) \! \int \! dp \,\, \theta(p_0) \delta(p^2 - m^2) \,
 \mbox{Si}(p_0/2n)  \, \overline{\widetilde{h}(p/n^2)} \, 
\widetilde{g} (p) \, , \\
& \langle h_{2,n} | g \rangle =
(2 \pi / n^{d-2} )  \! \int \! dp \,\, \theta(p_0) \delta(p^2 - m^2) \,
\overline{\widetilde{h}(p/n^2)} \, \widetilde{g} (p) \, , \\  
\end{split}
\end{equation*} 
using in $d=2$ dimensions once more the fact that  
$\widetilde{h}(0) = 0$. 
\end{proof}

In a last preparatory step we show that the sequences 
$\{ f_{1,n} , f_{2,n} \}_{n \in \NN}$ in the
preceding lemma can be used to construct sequences 
$\{ g_{1,n} , g_{2,n} \}_{n \in \NN}$ such that the resulting 
pairs satisfy the condition in Lemma {\ref{3.1}}.

\begin{lemma} \label{3.3}
Let  $\{ f_{1, n}, f_{2, n} \}_{n \in \NN}$ be sequences of 
test functions as in the preceding lemma. There exists
a dense set of translations $a \in \RR^d$ such that for any  
given $a$ in this set the corresponding sequences 
$\{ \tau_a f_{1, n}, \, \tau_a f_{2, n} \}_{n \in \NN}$,
where $(\tau_a f) (x) \doteq f(x-a)$, satisfy
$$ \langle f_{1, n} + i  f_{2, n} | \tau_a (f_{1, n} -i f_{2, n})
\rangle \neq 0 \, , \quad n \in \NN \, . $$
\end{lemma}

\begin{proof} 
Recalling that $M$ denotes the mass shell, one has 
$\widetilde{(f_{1,n} \pm i f_{2,n})} \upharpoonright M \neq 0$, 
$n \in \NN$. In fact, since
$f_{1,n} \pm i f_{2,n}$ has compact support, its Fourier 
transform is entire analytic and can vanish on
the mass shell only on a closed set of measure $0$. 
Thus the wave functions 
$\widetilde{\tau_a(f_{1,n} - i f_{2,n})} \upharpoonright M$, $a \in \RR^d$, 
form a total set in $\Hs_1$ for each $n \in \NN$. Next, 
observe that each function $a \mapsto  
\langle f_{1, n} + i  f_{2, n} | \tau_a (f_{1, n} -i f_{2, n}) \rangle
$, $n \in \NN$, extends to an analytic function in the forward tube
$\RR^d + i V_+$ as a consequence of the relativistic spectrum condition.
Hence, by the Edge--of--the--Wedge Theorem and the preceding
remarks, it can vanish only on a closed, nowhere dense subset
$S_n \subset \RR^d$ for each $n \in \NN$. Thus
$\bigcup_{n \in \NN} S_n$ is a meager set in the Baire space $\RR^d$. 
Its complement is therefore dense, proving the statement.  
\end{proof}

We are now in a position to prove that the net $\Rs$ defined in this  
model is maximally nonlocal. As a matter of fact, the following
somewhat stronger statement holds.

\begin{prop} \label{3.4}
Let $\Os_1$ and $\Os_2$ be spacelike separated regions 
in $\!$Minkowski space. 
Then $C(\Rs_e (\Os_1), \Rs_e (\Os_2)) = 1$,
where $\Rs_e$ denotes the subnet generated by all even polynomials 
in the smeared field $\phi$. 
\end{prop}

\begin{proof}
Because of the Poincar\'e covariance of the net, one may assume 
that $\Os_1$ contains the point $0$ in its interior. 
Choosing some suitable $a \in \RR^d$, it then follows from the preceding
Lemma that for sufficiently large  $n \in \NN$,
the supports of the functions $f_{1,n}, f_{2,n}$ 
constructed in Lemma \ref{3.1} are  
contained in $\Os_1$, those of $\tau_a f_{1,n}, \tau_a f_{2,n}$ in $\Os_2$, 
and $ \langle f_{1, n} + i  f_{2, n} | \tau_a (f_{1, n} -i f_{2, n})
\rangle \neq 0$ for all $n \in \NN$. In view of the support
properties of $f_{1,n}, f_{2,n}$, the 
projections $P_n = \frac{1}{2}(\idty + i \phi(f_{1,n}) \phi(f_{2,n}))$
are contained in $\Rs_e(\Os_1)$  and the projections 
 $Q_n = \frac{1}{2}(\idty + i \phi(\tau_a f_{1,n}) \phi(\tau_a
 f_{2,n}))$ in  $\Rs_e(\Os_2)$, provided $n \in \NN$ is sufficiently 
large. Moreover, as a consequence of Lemma \ref{3.1} and the fact that
the above scalar products are different from $0$, one has 
$P_n \wedge Q_n = 0$,  $n \in \NN$. 

Bearing in mind  that the scalar products of the 
one--particle states are invariant under translations, the
vacuum expectation values of $P_n$ and $Q_n$ are given by 
$ \langle \Omega, P_n \Omega \rangle = \langle \Omega, Q_n \Omega \rangle
= \frac{1}{2} (1 + i \langle f_{1,n} | f_{2,n} \rangle )$.  
Thus these expectation values 
converge to $1$ in the limit of large $n$ according to 
Lemma \ref{3.2}, and consequently $P_n \, \Omega \rightarrow \Omega$,
$Q_n \, \Omega \rightarrow \Omega$ strongly. Moreover, 
\begin{equation*}
\begin{split}
&  \sup_{E,F} \, 
|\langle \Omega, E \wedge F  \,  \Omega \rangle - 
\langle \Omega, E  \,  \Omega \rangle \, \langle  \Omega ,  F  \,  \Omega
\rangle | \\
& \geq \lim_{n \rightarrow \infty} \,
|\langle \Omega, P_n \wedge Q_n  \,  \Omega \rangle - 
 \langle  \Omega, P_n  \,  \Omega \rangle \, 
\langle  \Omega, Q_n  \, \Omega \rangle |  = 1 \, ,
\end{split}
\end{equation*}
where the supremum extends over all projections $E \in \Rs_e(\Os_1)$
and  $F \in \Rs_e(\Os_2)$.
In order to see that this lower bound holds for all normal
states, one makes use of the fact that the projections $P_n, Q_n$ form
central sequences. To verify this, note that for each real test function
$g$ one obtains by a straightforward application of the
anticommutation relations the bound
$$ \| \, [ P_n ,  \phi(g) ] \, \| \leq 
|\langle f_{1,n} | g \rangle| \, \| \phi(f_{2,n}) \| +
|\langle f_{2,n} | g \rangle| \, \| \phi(f_{1,n}) \| \, .
$$
The right hand side of this inequality tends to $0$ for large $n$
according to Lemma~\ref{3.2}. Since $P_n \Omega \rightarrow \Omega$,
strongly, and  $\Omega$ is cyclic for the algebra of all polynomials 
in the smeared fields, it follows that 
$P_n \rightarrow 1$ in the strong operator topology on $\Hs$. By
the same reasoning one sees that $Q_n \rightarrow 1$ as well.
Hence, recalling that $P_n \wedge Q_n = 0$, 
one obtains for any normal state $\omega$  
$$ \sup_{E,F} \, |\omega (E\wedge F) - \omega (E) \omega (F)|
\geq \lim_{n \rightarrow \infty} \,
|\omega (P_n \wedge Q_n) - \omega (P_n) \omega (Q_n)| = 1 \, .
$$
Since, from the outset,   
$C(\Rs_e (\Os_1), \Rs_e (\Os_2)) \leq 1$, the statement follows.
\end{proof}

Another expression of the nonlocality  
of the present model is the absence of any 
nontrivial operator which commutes with all operators in 
the spacelike complement of its localization region. We exhibit
this fact in the following Proposition.

\begin{prop}  \label{3.5}
Let $W \in \Ws$ be any wedge region. Then
$\Rs(W)^\prime \cap \Rs(W^\prime) = \CC \cdot 1$. 
\end{prop}

\begin{proof} It follows from Lemma \ref{2.1} that 
$\Rs (W^\prime) \subset \widehat{\Rs} (W)^\prime$ and 
consequently $\Rs(W)^\prime \cap \Rs(W^\prime)
\subset \Rs(W)^\prime \cap \widehat{\Rs}(W)^\prime \doteq \Is(W) $. 
The $*$--algebra $ \Is(W)$ is invariant under the adjoint action
of the unitary involutions $V$, $Z$, as well as of the unitaries representing
the boosts which leave the wedge $W$ invariant. 
Let $E \in  \Is(W)$ be any even element under the adjoint action of
$Z$, \ie $EZ = EZ$, and let $f$ be any test function with support
in $W$. It then follows from the results in Sect.\ 2 that
\begin{equation*}
\begin{split}
& 0 = [ E, \phi(f) ] = [ E, \phi_+(f) + \phi_-(f) ] \, , \\
& 0 = [ V E V, \phi(f) ] = 
V \, [E, ( \phi_+(f) - \phi_-(f)) Z ]\,  V =
V \,  [E, ( \phi_+(f) - \phi_-(f)) ] \, Z V \, ,
\end{split}
\end{equation*}
and consequently $[ E, \phi_\pm (f) ] = 0$ whenever $\mbox{supp}\, f
\subset W$. As the operator valued functions 
$ \RR^d \ni a \mapsto  [ E, \phi_\pm (\tau_a f) ] $ 
can be continued analytically 
into the forward and backward tubes, respectively, and the 
translated test functions $\tau_a f$ have support in $W$ for 
an open set of translations $a \in \RR^d$,
it follows from the Edge-of-the-Wedge Theorem that 
 $[ E, \phi_\pm (f) ] = 0$ for all test functions $f$. 
Since the creation and
annihilation operators act irreducibly on 
the Fock space $\Hs$, 
$E = c \cdot 1$ for some $c \in \CC$ .

Next, let $O_1, O_2 \in \Is (W)$ be two odd elements under the
adjoint action of $Z$, \ie $O_j Z = - Z O_j$, $j = 1,2$. Then
$O_1 O_2$ is even and consequently $O_1 O_2 = c \cdot 1$ for some
$c \in \CC$. Let $v(t), t \in \RR$, be the one--parameter 
group of boosts leaving the wedge $W$ invariant. 
Since $Z$ commutes with all Poincar\'e transformations, one
may replace in the preceding equation the operator $O_2$
by $O_2(t) \doteq U(v(t)) O_2 U(v(t))^{-1}$, giving 
$O_1 O_2(t) = c(t) \cdot 1$, and consequently 
$$  
O_1  \, U(v(t))  \,  O_2 O_2^* \Omega = 
O_1 \, O_2(t) O_2^*(t) \Omega 
= c(t) \,  O_2^*(t) \Omega = c(t) \,  U(v(t)) \,  O_2^* \Omega \, , 
\quad t \in \RR \, .
$$ 
Now $\mbox{w--lim}_{\, t \rightarrow \infty} \,  U(v(t))  \Psi = \langle \Omega,
\Psi \rangle \, \Omega$ for any $\Psi \in \Hs$, so the preceding
equality implies $ \langle  \Omega,  O_2 O_2^*  \, \Omega \rangle \, O_1
\Omega = c(\infty)  \,  \langle  \Omega,  O_2^*  \, \Omega \rangle \, \Omega
$, where $c(\infty) \doteq \lim_{t \rightarrow \infty} c(t)$ must exist,
since the other limits in the above equation exist. 
Since $O_2$ is odd and $\Omega$ is invariant under the action
of $Z$, one has $\langle  \Omega,  O_2^*  \, \Omega \rangle = 0$,
so either $O_1 \Omega = 0$ or $ O_2^* \Omega = 0$.  Since $\Omega$
is separating for $\Is (W)$ (it is cyclic for $\Rs (W)$, as can be proven
by standard arguments \cite{StWi,Jost,Ar1}), one concludes that 
there are no nonzero odd elements in $\Is (W)$. But any element of $\Is (W)$
can be decomposed into the sum of an even and an odd one under
the action of $Z$, so  $\Is (W) = \CC \cdot 1$, proving the assertion 
for the net $\Rs$.
\end{proof}

     Although the present model is nonlocal in a very strong sense, 
we nonetheless want to uncover in the subsequent 
sections some interesting   
properties which are consistent with Einstein causality. 
As we shall see, the model has many features which one normally attributes 
to local theories. 

\section{Modular structure and weak locality}

\setcounter{equation}{0}

     We determine in this section the modular groups associated with
the vacuum vector $\Omega$ and the algebras $\Rs(W)$ corresponding to
wedge regions $W \in \Ws$.  It will turn out that they coincide with
those found in local theories by Bisognano and Wichmann \cite{BiWi,
BiWi2}; however, the modular conjugations differ from those of local
theories. The results will allow us to establish a weak form 
of locality of wedge algebras.

We begin by noting that, within the framework given above, standard arguments
\cite{StWi,Jost,Ar1} entail that both nets $\net,\hatnet$ are irreducible
and that $\Omega$ is cyclic for both $\Rs(\Os)$ and $\widehat{\Rs}(\Os)$,
for any nonempty open $\Os$. Lemma \ref{2.1} then implies that
$\Omega$ is also separating for  $\Rs(\Os)$ and $\widehat{\Rs}(\Os)$,
whenever $\Os^\prime$ is open and nonempty. Thus, in particular,  
$\Omega$ is cyclic and separating for all wedge algebras 
$\Rs(W),\widehat{\Rs}(W)$, $W \in \Ws$. The Tomita-Takesaki modular theory, 
cf.\ \cite{BratRob1,KadRing}, is therefore applicable to the pairs 
$(\Rs(W),\Omega)$, $(\widehat{\Rs}(W),\Omega)$, for all wedges $W$.

It is our aim to compute the corresponding modular objects. This task
is greatly facilitated by making use of known results 
\cite{Jost,BiWi,BrGuLo}. Let $v_R(t)$, $t \in \RR$, be the  
boost subgroup in $\Pid$ which induces a positive timelike 
flow on the wedge $W_R$ and which is periodic for 
imaginary $t$ with period $2 \pi$. One then has 
$$ v_R(i\pi) = \theta_R \, , $$
where $\theta_R \in \Ps_+$ (the proper Poincar\'e group) 
is the reflection about the edge of $W_R$. Now if 
$f \in \Ss(\RR^d)$ has support in $W_R$, then $|f\rangle$ is
known to lie in the domain of the positive self-adjoint operator 
$U(v_R(i\pi))$ and  
$$ U(v_R(i\pi) | f \rangle = | {\theta_R f} \rangle \, , $$
where $(\theta_R f)(x) \doteq f(\theta_R x)$. Moreover, 
the antilinear map $| f \rangle \mapsto 
| \theta_R \overline{f} \rangle $ is isometric.
Using the functorial character of the representation $U$, these
statements can be extended to the many particle states. One has    
$$
U(v_R(i \pi)) \, | f_1 \rangle \wedge \cdots \wedge | f_n \rangle =
|\theta_R f_{1} \rangle \wedge \cdots \wedge | \theta_R f_{n} \rangle 
\, ,
$$
provided all functions $f_1, \dots , f_n$ have support in $W_R$;
the operator $U(\theta_R)$ fixed~by
\begin{equation} \label{tcp}
U(\theta_R) \, | f_1 \rangle \wedge \cdots \wedge | f_n \rangle \doteq
|\theta_R \overline{f_{1}} \rangle \wedge \cdots \wedge 
| \theta_R \overline{f_{n}} \rangle 
\end{equation}
is an antiunitary involution extending the representation $U$ of $\Pid$ to
a representation of $\Ps_+$. (This statement amounts to the 
PCT--theorem in even dimensions $d$.) Taking into account the 
antisymmetry properties of the many particle states, one has 
$$
|f_{n} \rangle \wedge \cdots \wedge 
| f_{1} \rangle = 
(-1)^{n(n-1)/2} \, | f_1 \rangle \wedge \cdots \wedge | f_n \rangle =
V \, | f_1 \rangle \wedge \cdots \wedge | f_n \rangle \, ,
$$  
where $V$ is the unitary operator introduced in Section 2.\
Combining the above relations, one arrives at
$$
V U(\theta_R) \, U(v_R(i \pi)) \, 
| f_1 \rangle \wedge \cdots \wedge | f_n \rangle =
|\overline{f_{n}} \rangle \wedge \cdots \wedge 
|\overline{f_{1}} \rangle \, ,
$$
provided $f_1, \dots , f_n$ have the support properties 
stated above.

It is straightforward to restate these results in terms of the field operators.
Making repeated use of relation (\ref{fieldaction}), one gets
$$
\phi(f_1) \cdots \phi(f_n) \Omega =
\sum_p (-1)^{\tau(p)} \, \langle \overline{f_{p_1}} | f_{p_2} \rangle \cdots 
\langle \overline{f_{p_{k-1}}} | f_{p_k} \rangle \cdot |f_1 \rangle \cdots 
\overset{p_1}{\vee} \cdots \overset{p_k}{\vee} \cdots |f_n \rangle \, ,
$$
where the sum extends over all ordered pairs $p \subset \{1, \dots ,n \}$ 
and $\tau(p)$ is the number of transpositions needed to transform
$1, \dots, n$ into $p_1 , \dots , p_k, 1, \dots \overset{p_1}{\vee} \dots 
 \overset{p_k}{\vee} \dots , n$. Noticing that this number is equal
(modulo $2$) to the number of transpositions needed to transform 
$n, \dots ,1$ into  $p_k , \dots , p_1, n, \dots \overset{p_k}{\vee} \dots 
 \overset{p_1}{\vee} \dots , 1$ and bearing in mind that $U(\theta_R)$ is 
antilinear, one obtains with $f_1, \dots , f_n$ as above
\begin{equation*}
\begin{split}
& V U(\theta_R) \, U(v_R(i \pi)) \, \phi(f_1) \cdots \phi(f_n) \Omega \\
& = \sum_p (-1)^{\tau(p)} \, \langle \overline{\overline{f_{p_k}}} | \overline{f_{p_{k-1}}} \rangle \cdots 
\langle \overline{\overline{f_{p_2}}} | \overline{f_{p_1}} \rangle \cdot 
|\overline{f_n} \rangle \cdots 
\overset{p_k}{\vee} \cdots \overset{p_1}{\vee} \cdots |\overline{f_1} \rangle \\
& = \phi(\overline{f_n}) \cdots \phi(\overline{f_1}) \Omega 
= \big( \phi(f_1) \cdots \phi(f_n) \big)^* \Omega \, .
\end{split}
\end{equation*}
Since the polynomials in the fields $\phi(f)$ with 
$\mbox{supp} \, f \subset W_R$ are weakly dense in $\Rs(W_R)$ and
generate from the vacuum vector $\Omega$ a domain of essential
self-adjointness for $U(v_R(i\pi))$, it follows that 
$S_R = V U(\theta_R) \, U(v_R(i \pi))$ is the Tomita conjugation
for the pair $(\Rs(W_R), \Omega)$. As the expression given for
$S_R$ is already its polar decomposition, it is apparent that
$\Delta_R = U(v_R(2i \pi))$ is the modular operator and 
$J_R = V U(\theta_R)$ the modular conjugation associated with
$(\Rs(W_R), \Omega)$. Making use of relations (\ref{fieldaction})
and (\ref{tcp}), one checks that 
$J_R \phi(f) J_R = V \phi(\theta_R \overline{f}) V = 
\widehat{\phi}(\theta_R \overline{f})$. Thus
$\Rs(W_R)^\prime = J_R \Rs(W_R) J_R = \widehat{\Rs} ({W_R}^\prime)$,
where the first equality follows from Tomita--Takesaki theory 
and the second one from the geometrical fact that 
$\theta_R W_R = W_L = {W_R}^\prime$. Since the Poincar\'e 
transformations commute with $V$, the modular objects 
corresponding to $(\widehat{\Rs}(W_R), \Omega)$ coincide
with those of $(\Rs (W_R), \Omega)$, and one has 
$\widehat{\Rs}(W_R)^\prime = 
J_R \widehat{\Rs}(W_R) J_R = \Rs ({W_R}^\prime)$. 
We summarize these results in the following proposition.  

\begin{prop} \label{4.1}
The modular operator and conjugation corresponding to the 
pair $(\Rs (W_R), \Omega)$ are given by $\Delta_R = U(v_R(2i \pi))$
and $J_R = V U(\theta_R)$, respectively. Moreover,
\begin{equation} 
\Rs(W_R)^\prime = \widehat{\Rs}({W_R}^\prime) \, .
\end{equation}
These statements hold likewise for $(\widehat{\Rs} (W_R), \Omega)$,
if one interchanges $\Rs$ and $\widehat{\Rs}$ in the preceding equality.
\end{prop}

Since the proper Poincar\'e group $\Ps_+$ acts covariantly and
transitively on the wedge algebras, this result extends
to the pairs $(\Rs(W), \Omega)$ for arbitrary wedge regions $W$
in an obvious manner. In particular, the corresponding modular groups
are induced by the boosts leaving the respective wedge $W$
invariant. This property of a net has come to be called
Modular Covariance, cf.\ the review article \cite{Bor}.
It readily follows that this model satisfies the superstability
conditions studied in \cite{AdS}.
     However, since the net is not local, the modular conjugations do not
act in a geometric manner, \ie the Condition of
Geometric Modular Action \cite{BDFS} is not satisfied in this model.
Nonetheless, the modular structure exhibited above allows
one to establish a rudiment of locality which 
{\em prima facie} is stronger than the property of
``weak locality'' established for the present model in \cite{Jost}.

\begin{prop} \label{4.2}
Let $W$ be any wedge region. Then
$$ 
\langle \Omega, A B \Omega \rangle = \langle \Omega, B A \Omega
\rangle 
\quad \mbox{for} \
A \in \Rs(W), B \in \Rs(W^\prime) \, .
$$
\end{prop}

\begin{proof} 
It suffices to prove this statement for the 
wedge $W_R$. Since $\Rs({W_R}^\prime) = \widehat{\Rs} (W_R)^\prime$,
the modular objects corresponding to $(\Rs({W_R}^\prime), \Omega)$
are given by $\Delta_R^{-1}, J_R$. So one has
\begin{equation*}
\begin{split}
& \langle \Omega, A B \, \Omega \rangle = \langle A^* \Omega, B \Omega
\rangle = 
\langle J_R \Delta_R^{1/2} A \Omega, J_R \Delta_R^{-1/2} B^* \Omega \rangle \\
& = \langle \Delta_R^{-1/2} B^* \Omega, \, \Delta_R^{1/2} A \Omega \rangle = 
\langle \Omega, B A \Omega \rangle \, ,
\end{split}
\end{equation*}
as claimed.
\end{proof}

\section{Independence properties}

\setcounter{equation}{0}

     The preceding information on the modular operators 
corresponding to wedge algebras and the vacuum state
allows one to establish strong independence properties of 
pairs of such algebras associated with spacelike separated
wedge regions. Our first result says that any pair of such algebras 
has no nontrivial operator in common. This fact is, 
in a sense, complementary to the statement of 
Proposition \ref{3.5}. The proof is based on standard
arguments, cf.\  \cite{AdS}, which we recall here for the 
convenience of the reader.  

\begin{prop} \label{5.1}
For any wedge $W \in \Ws$, one has 
$\Rs(W) \cap \Rs(W^\prime) = \CC \cdot 1$.
\end{prop}

\begin{proof}
Because of covariance, it suffices to prove the statement
for the wedge $W_R$. Now according to Proposition \ref{4.1},
$J_R, \Delta_R$ are the modular objects corresponding to 
$(\Rs(W_R), \Omega)$, and $J_R, \Delta_R^{-1}$ are those
corresponding to $(\Rs(W_R^\prime) = \widehat{\Rs}(W_R)^\prime, \Omega)$.
Thus for any  $A \in \Rs(W_R) \cap \Rs(W_R^\prime)$ one has
$J_R \Delta_R^{1/2} A \Omega = A^* \Omega = 
J_R \Delta_R^{-1/2} A \Omega$,
and consequently $\Delta_R A \Omega = A \Omega$. But this implies 
$ A \Omega = \Delta_R^{it} A \Omega = U(v_R(2 \pi t)) A \Omega$, $t \in \RR$.
Proceding in this equality to the limit $t \rightarrow \infty$,
one obtains $ A \, \Omega = \langle \Omega, A \, \Omega \rangle \,
\Omega$. Since $\Omega$ is separating for $\Rs(W_R) \cap \Rs(W_R^\prime)$,  
the assertion follows.
\end{proof}

For strictly spacelike separated wedges $W_1, W_2 \in \Ws$, \ie 
$\overline{W_1} \subset W_2{}^\prime$, one can establish a 
substantially stronger variant of this result, which expresses  
the algebraic independence of the corresponding wedge algebras.
In local quantum field theory, an analogous result was 
proven by Schlieder \cite{Sch} and Roos \cite{Ro}, 
using quite different arguments. 

\begin{prop} \label{5.2}
Let $W_1,W_2 \in \Ws$ be strictly spacelike separated. 
For any $n \in \IN$, $A_{1,k} \in \Rs(W_1)$ and
$A_{2,k} \in \Rs(W_2)$, $k = 1,\ldots,n$, such that
$\sum_{k=1}^n A_{1,k} \, A_{2,k} = 0$, one must have
$$\sum_{k=1}^n \psi(A_{1,k}) A_{2,k} = 0 = \sum_{k=1}^n A_{1,k} \, 
\psi (A_{2,k}) \, ,  $$
for all normal linear functionals $\psi$ on $\Bs(\Hs)$. In particular, if 
$A_1 A_2 = 0$, then either $A_1 = 0$ or $A_2 = 0$.
\end{prop}

\begin{proof}
Since the proof follows in the steps of the arguments given in 
\cite[Sect.\ III]{AdS}, only the necessary changes to be made will be
indicated.  

Due to Poincar\'e covariance, there is no loss of generality to 
assume that $\overline{W_1} \subset W_R$ and 
$\overline{W_2} \subset W_R^\prime$. 
Let $\Ps_R \subset \Pid$ be the subgroup generated by the 
boosts $v_R(\RR)$, which leave the wedge $W_R$ invariant, 
and the two--dimensional subspace of 
spacetime translations which are orthogonal to the 
edge of  $W_R$; thus $\Ps_R$ is isomorphic to the identity 
component of the Poincar\'e group in two dimensions. The
following facts enter into the proof: (a) There is an open 
neighborhood $\Ns_R \subset \Ps_R$ of the identity such that 
${W_1} \subset \lambda_0 \lambda_1 W_R$ and 
${W_2} \subset \lambda_0 \lambda_1 W_R^\prime$
for all $\lambda_0,  \lambda_1 \in \Ns_R$. 
(b) The group generated by $\{ \lambda v_R(t) \lambda^{-1} : t \in \RR, \,
\lambda \in  \Ns_R \}$ coincides with $\Ps_R$.
(c) Making use of Proposition (\ref{4.1}) and the
irreducibility of the net $\widehat{\Rs}$, one has 
\begin{equation} \label{generate}
\bigvee_{\lambda \in \Ps_R} U(\lambda) \Rs(W_R)^\prime \, U(\lambda)^{-1} =
\bigvee_{\lambda \in \Ps_R} U(\lambda) 
\widehat{\Rs}(W_R^\prime) U(\lambda)^{-1} =
\Bs(\Hs) \, . 
\end{equation}    
Taking into account  that $t \mapsto U(v_R(2 \pi t))$ and
 $t \mapsto U(v_R(- 2 \pi t))$ are the 
modular groups  corresponding to $(\Rs(W_R), \Omega)$ and 
 $(\Rs(W_R)^\prime, \Omega)$, respectively, 
the necessary ingredients for the
arguments given in \cite{AdS} are therefore in place
here, as well. More specifically, the statements and proofs of 
 \cite[Lemma 3.1, Lemma 3.2]{AdS} carry over to the present 
situation, if one replaces the group 
$SO(2,n-1)$ by  $\Ps_R$. The assertion then follows as
in \cite[ Proposition 3.3]{AdS}.
\end{proof}

\section{Coherent families of local observables}  
\label{sec6}

\setcounter{equation}{0}

     As nonlocal as we have seen the net $\wrnet$ to be, nonetheless it
accomodates quantities which one may assign to point--, string-- or
brane--shaped regions and which commute when spatially separated.
In this section we shall exhibit explicit examples of such operators. 

     Let $W_1, W_2 \in \Ws$ be wedges such that $\overline{W}_2 \subset W_1$. 
Because of Poincar\'e covariance, we may assume that $W_1 = W_R$ 
and $W_2 = W_R + (a,0, \dots 0)$ for some $a > 0$. So 
both edges of these wedges lie in the time--$0$--plane.
We want to show that there exist nontrivial field operators 
$\phi(h) \in \Rs(W_1)$ which anticommute with all field
operators $\phi(f) \in \Rs(W_2)$. To this end, we need the following
preparatory lemma.

\begin{lemma} \label{6.1}
Let $a > 0$. 
There exist test functions $\widetilde{k} \in \Ss (\RR^{d-1})$ such that   
\begin{equation*}
\bx \mapsto k(\bx) = (2 \pi)^{(1-d)/2} 
\int \! d \bp  \,\, \widetilde{k} (\bp) \,\, e^{i \, \sbp  \sbx} 
\end{equation*}
vanishes in the region $\{ \bx \in \RR^{d-1} :  x_1 < 0 \}$ and 
\begin{equation*}
\bx \mapsto  (2 \pi)^{(1-d)/2}
\int \! \frac{d \sbp}{\sqrt{\sbp^2 + m^2}} \,\, \widetilde{k}
(\bp) \,\, e^{i \, \sbp \sbx } 
\end{equation*}
vanishes in the region $\{ \bx \in \RR^{d-1} :  x_1 > a \}$, respectively.
\end{lemma}

\begin{proof} Begin by noting that for fixed $\kappa > 0$ the distributions 
$$ 
y \mapsto \int \! dq \, \sqrt{q - i \kappa} \,  e^{i qy}
\quad  \mbox{and} \quad 
y \mapsto \int \! dq \, \frac{1}{\sqrt{q + i \kappa}} \,  e^{i qy} 
$$
vanish on $\RR_-$ and $\RR_+$, respectively, because of the
analyticity and temperedness properties of the integrands. 
Consider now, for 
given $l \in \Ss (\RR^{d-1})$ 
with compact support in $ \{ \bx \in \RR^{d-1} : 0 < x_1 < a \} $,
the functions 
$$ 
\widetilde{k} (\bp) \doteq 
\sqrt{p_1 - i \sqrt{\sbp_\perp^2 + m^2}} \
\widetilde{l} (\bp) \, , \quad
 \frac{1}{\sqrt{\sbp^2 + m^2}} \, \widetilde{k} (\bp) =
 \frac{\widetilde{l} (\bp)}{\sqrt{p_1 + i \sqrt{\sbp_\perp^2 + m^2}}} \, ,
$$
where $\bp_\perp$ is the component of $\bp$ which is orthogonal to 
the 1--direction. Since the Fourier transform of a product of 
functions coincides with the convolution of the  
Fourier transforms of the individual factors, the assertion follows.
\end{proof}

Since $\| \phi(g) \|^2 \leq \langle g | g \rangle + 
\langle \overline{g} | \overline{g} \rangle$, the field   
$\phi$ can be extended by continuity to (generalized) functions of the form 
$x \mapsto h(x) = \delta(x_0) \, k(\bx)$, where $k$ is any test 
function as in the preceding lemma. Because of the 
support properties of $k$ and the fact that $\Rs(W_1)$
is norm--closed, it follows that $ \phi(h) \in \Rs(W_1)$. Now 
the function 
$$ x \mapsto \{ \phi(h),\phi (x)\} = (2 \pi)^{(1-d)/2} 
\int \!  \frac{d \sbp}{\sqrt{\sbp^2 + m^2}} \, \tilde{k}(\bp) \, 
\cos (x_0 \sqrt{\bp^2 + m^2}) \, e^{i \sbp \sbx} \cdot \idty \, $$
is a solution of the Klein--Gordon equation whose Cauchy data
at time $x_0 = 0$ vanish in the region 
$\{ \bx \in \RR^{d-1} :  x_1 > a \}$ according to the preceding
lemma. Because of the hyperbolic nature of the Klein--Gordon equation,
this function thus vanishes also in the causal completion of that
region, \ie in the wedge $W_2$. Thus we conclude that 
$\{ \phi(h), \phi(f) \} = 0$ for all test functions $f$ with
$\mbox{supp} \, f \subset W_2$. 

We finally note that the Fourier transforms of the functions
$h$ defined above can vanish on the mass shell $M$
only on subsets of measure $0$, since the functions $l$
entering in the preceding
lemma in the construction of $k$ have compact supports.
As a consequence, $\Omega$ is cyclic for the algebra generated
by the field operators $\{ \phi(\tau_a h) : a \in N \}$, 
where $N \subset \RR^d$ is any open neighborhood of the origin. 
So we have established the following result.

\begin{lemma}  \label{lemma6.2}
Let $W_1,W_2 \in \Ws$  be wedges such that $\overline{W_2} \subset W_1$.
Then there exist nonzero elements $\phi(h) \in \Rs(W_1)$ such that
\begin{equation} 
\{ \phi(h) , \phi(f) \} = 0 \quad \text{whenever} \ \,  
\text{\rm supp} f \subset W_2 \, .
\end{equation}
Indeed, there are so many such elements that $\Omega$ is a cyclic
vector for the algebra they generate.
\end{lemma}

We make use of this result in order to define nontrivial algebras of local 
operators, which can be assigned to the intersection of certain
specific wedge regions.
We begin with the following definition. \\[3mm]
\textbf{Definition} \ Let $W_0 \in \Ws$ be a fixed wedge. 
(a) A wedge  $W \in \Ws$ 
is said to be coherent with $W_0$ if there exists some translation
$a \in \RR^d$ such that $W + a \subset W_0$~; the set of all wedges
which are coherent with  $W_0$ is denoted by $\Ws_0$. (b) The
subgroup of $\Pid$ generated by the translations $\RR^d$ and the 
stability group of $W_0$ in $\Pid$ is denoted by $\Ps_0$;
it is the largest subgroup of $\Pid$ whose 
action leaves the set  $\Ws_0$  of coherent wedges invariant. 

Fixing a wedge $W_0$ and corresponding coherent
family $\Ws_0$ as above, we assign to 
each pair of wedges $W_1, W_2 \in \Ws_0$
such that $\overline{W_2} \subset W_1$, \ie 
$W_1 \cap W_2^\prime$ has nonempty interior, the algebra 
\begin{equation} \label{6.2}
\As_0 (W_1 \cap W_2^\prime) \doteq \Rs(W_1) \cap \Rs(W_2)^\prime \, .
\end{equation}
As an immediate consequence of the preceding lemma, all even polynomials 
of the operators $\phi(h)$ described there are elements of this 
algebra, so it is clearly nontrivial. If  $W_3, W_4 \in \Ws_0$ are
such that $W_1 \cap W_2^\prime \subset W_3 \cap W_4^\prime$, it follows
after a moment's reflection that $W_1 \subset W_3$ and  
$W_4 \subset W_2$. Hence 
$$
\As_0 (W_1 \cap W_2^\prime) = \Rs(W_1) \cap \Rs(W_2)^\prime
\subset \Rs(W_3) \cap \Rs(W_4)^\prime  = 
\As_0 (W_3 \cap W_4^\prime) 
$$
since the net $\Rs$ is inclusion preserving.
If, on the other hand, the regions $W_1 \cap W_2^\prime$ and 
$ W_3 \cap W_4^\prime $ are spacelike separated, 
then either  $W_1 \subset W_4$ or $W_3 \subset W_2$. Hence, 
in either case,
$$
\As_0 (W_1 \cap W_2^\prime) =
\Rs(W_1) \cap \Rs(W_2)^\prime \subset 
\Rs(W_4) \vee \Rs(W_3)^\prime  
= \As_0 (W_3 \cap W_4^\prime)^\prime \, . 
$$
So the map $W_1 \cap W_2^\prime \mapsto \As_0 (W_1 \cap W_2^\prime)$
defines a local net based on intersections of coherent wedges. 
Since these intersections are infinitely extended in $(d - 2)$ spatial 
directions and arbitrarily thin in the remaining $2$ directions, the
underlying operators may be thought of as being point--, string-- 
or brane--localized, respectively, if $d=2,3$ or $\geq 4$. 
In view of the covariant action 
of the Poincar\'e transformations on the net $\Rs$, one also has
$$
U(\lambda_0) \As_0 (W_1 \cap W_2^\prime) U(\lambda_0)^{-1} 
= \Rs(\lambda_0 W_1) \cap \Rs(\lambda_0 W_2)^\prime = 
\As_0 (\lambda_0 (W_1 \cap W_2^\prime) ) \, , \quad \lambda_0 \in
\Ps_0 \, ,
$$
since $\lambda_0 W_1, \lambda_0 W_2 \in \Ws_0$.
We summarize these results.

\begin{prop} Let $\Ws_0$ be a coherent set of wedges.
The corresponding net  $\As_0$, defined in (\ref{6.2}), is 
local, $\Ps_0$--covariant and nontrivial.
\end{prop}

Thus, in spite of the nonlocality of the net $\Rs$, its 
net structure can be used to identify 
coherent families of subalgebras, which may be 
regarded as theories of spatially extended local operators.
Knowing that these subalgebras are nontrivial, it is natural to ask
how big they actually are. It is an interesting fact 
that the answer depends on the number $d$ of spacetime dimensions.
In the formulation of the subsequent results there appear the 
subspace $\Hs_e \subset \Hs$ of states with an even particle 
number and the nets $\Rs_e$, $\widehat{\Rs}_e$ generated by 
even polynomials in the fields $\phi$ and $\widehat{\phi}$, respectively.

\begin{prop} \label{prop6.4}
Let $d \geq 3$, let $\Ws_0$ be the set of all wedges in 
$\RR^d$ which are coherent with a given wedge $W_0$,  
and let $\overline{W_2} \subset W_1 \in \Ws_0$.
Then \\[1.5mm]
(a) \ $\overline{\As_0(W_1 \cap W_2^\prime) \, \Omega} = \Hs_e$, \\[1.5mm]
(b) \ $ \bigvee_{\, W_1 \cap W_2^\prime \, \subset \, W_0} \, \As_0(W_1 \cap
W_2^\prime) = \Rs_e(W_0)$,  \\[1mm]
(c)  \ $ \bigvee_{\,  W_1 \cap W_2^\prime \, \subset \, W_0^\prime}
\, \As_0(W_1 \cap W_2^\prime) = \widehat{\Rs}_e(W_0^\prime)$. 
\end{prop} 
\begin{proof}
(a) Each algebra
$\As_0(W_1 \cap W_2^\prime)$ contains all even polynomials of the
field $\phi$ smeared with functions $h$ as in the statement 
of Lemma \ref{lemma6.2}. It follows that 
$ \Hs_e \subset \overline{\As_0(W_1 \cap W_2^\prime) \, \Omega}$. 
For the proof of the converse inclusion,  
note that $\As_0(W_1 \cap W_2^\prime)$ is stable under the 
adjoint action of $Z = (-1)^N$. Hence one must show that
there is no nonzero element $X \in \As_0(W_1 \cap W_2^\prime)$
satisfying $XZ = -ZX$, \ie being odd. 
To this end, one makes use of 
Lemma \ref{lemma6.2} once again and picks functions 
$h$ such that 
$\{ \phi(h) , \phi(f) \} = 0$ whenever $\text{\rm supp} \, f \subset W_1$.
As $\Rs(W_1)$ is generated by the operators $ \phi(f)$, it follows
that $\phi(h) Z \in \Rs(W_1)^\prime$, and since $X \in \Rs(W_1)$ is odd,  
$XZ$ must commute with all operators $\phi(h)$. On the other hand, 
$[X, \phi(g)] = 0$ whenever $\text{\rm supp} \, g \subset W_2$, since
$X \in \Rs(W_2)^\prime$. 

Now let $a$ be any translation 
along the edge of $W_0$ (such translations  exist if $d \geq 3$) and
let $B(a) \doteq U(a) B U(a)^{-1}$,  $B \in \Bs(\Hs)$.  
In view of the invariance of the wedges in  $\Ws_0$ under the 
action of $a$, one obtains for functions $h_1, \dots h_n$ and  
$g_1, \dots g_m$ as above the equality 
\begin{eqnarray} \label{6.3}
\lefteqn{\langle \phi(g_1) \cdots \phi(g_m)\Omega, \; (X^* XZ)(a) \;
\phi(h_1) \cdots \phi(h_n) \Omega \rangle} \nonumber \\
& & =  \langle X(a) \Omega, \big( \phi(g_m)^* \cdots \phi(g_1)^*
\phi(h_1) \cdots \phi(h_n) \big) \; (XZ)(a) \Omega \rangle \\ 
& & =  \langle X\Omega, \big( \phi(g_m)^* \cdots \phi(g_1)^*
\phi(h_1) \cdots \phi(h_n) \big) (-a) \; XZ \Omega \rangle \nonumber \, . 
\end{eqnarray}
Since the field $\phi$ is irreducible and 
the anticommutators $\{ \phi(u), \phi(v)(-a) \}$ 
vanish in norm for arbitrary test functions $u, v$ as $a$ tends to infinity, 
one obtains in this limit
$$ \big( \phi(g_m)^* \cdots \phi(g_1)^*
\phi(h_1) \cdots \phi(h_n) \big) (-a) \, \rightarrow \,
\langle \Omega, \phi(g_m)^* \cdots \phi(g_1)^*
\phi(h_1) \cdots \phi(h_n) \Omega\rangle \cdot \idty $$
in the weak operator topology. Combining these results
and taking into account that $Z \Omega = \Omega$, it follows that
$$ \lim_{a \rightarrow \infty} \, (X^* XZ)(a) = \langle \Omega, X^* X
\Omega \rangle \cdot 1 
$$
and, in a similar manner,  
$$ \lim_{a \rightarrow \infty} \, (X^* X)(a) = \langle \Omega, X^* X
\Omega \rangle \cdot 1 \, .$$
As $Z$ is invariant under the adjoint action of the 
translations, the latter two 
results are only compatible if $X \Omega =0$. But $\Omega$ is separating for
$\As_0(W_1 \cap W_2^\prime)$, hence $X=0$, proving that 
$\overline{\As_0(W_1 \cap W_2^\prime) \, \Omega} \subset \Hs_e $. \\[1mm]
(b) According to Proposition \ref{4.1}, the modular group
corresponding to the pair $(\Rs(W_0), \Omega)$ consists of
the unitary boost transformations leaving 
the wedge $W_0$ invariant. As the even subnet $\Rs_e$ is 
left invariant under their adjoint action, it follows that 
the modular group of  $(\Rs_e(W_0),
\Omega)$ coincides with the restriction of the
boosts to the subspace $\Hs_e$. 
Since these boosts act covariantly on the 
net $\As_0$, the algebra 
$ \bigvee_{\, W_1 \cap W_2^\prime \, \subset \, W_0} \, \As_0(W_1 \cap
W_2^\prime) \subset \Rs_e(W_0)$ is stable under their action. 
But $\Omega$ is cyclic 
in $\Hs_e$ for the algebras $ \As_0(W_1 \cap W_2^\prime)$, so one must 
have equality in this inclusion by a standard result 
in modular theory. Statement (c) 
follows from Proposition \ref{4.1} in a similar manner.
\end{proof}

     In a manner similar to the construction of the net $\As_0$, one 
can proceed from $\Rs$ to a $\Ps_0$--covariant 
field net $\Fs_0$ satisfying twisted locality (in analogy to 
theories of local Fermi fields). Introducing 
the twisted algebras 
$\Rs^t (\Os) \doteq \{ \phi(f) \, Z : \text{supp} \, f \subset
\Os \}^{\prime \prime}$, one defines for any
$\overline{W_2} \subset W_1 \in \Ws_0$
\begin{equation} \label{6.4}
\begin{split} 
& \Fs_0(W_1 \cap W_2^\prime) \doteq \Rs(W_1) \cap \Rs^t(W_2)^\prime \, , \\
& \Fs^t_0(W_1 \cap W_2^\prime) \doteq \Rs^t(W_1) \cap \Rs(W_2)^\prime \, .
\end{split}
\end{equation}
If the regions $W_1 \cap W_2^\prime$ and 
$ W_3 \cap W_4^\prime $ are spacelike separated, then 
$$
\Fs_0 (W_1 \cap W_2^\prime) =
\Rs(W_1) \cap \Rs^t(W_2)^\prime \subset 
\Rs(W_4) \vee \Rs^t(W_3)^\prime  
= \Fs^t_0 (W_3 \cap W_4^\prime)^\prime \, , 
$$
proving twisted locality. Since $Z$
commutes with the unitaries representing the Poincar\'e
transformations, it is also clear that the net $\Fs_0$
is $\Ps_0$--covariant. Moreover, 
it is an immediate consequence of Lemma \ref{lemma6.2} that
$\overline{\Fs_0(W_1 \cap W_2^\prime) \, \Omega} = \Hs$, so
$\Omega$ is cyclic for the field net. Finally, 
by the preceding proposition one has
for spacetime dimensions $d \geq 3$
$$\Rs(W_1) \cap \Rs(W_2)^\prime = \Rs_e(W_1) \cap \Rs(W_2)^\prime
= \Rs_e(W_1) \cap \Rs^t(W_2)^\prime \subset \Rs(W_1) \cap
\Rs^t(W_2)^\prime \, ,
$$
where the second equality follows from the fact that all 
elements of the net $\Rs_e$ commute with the unitary operator $Z$.
Hence  
$\As_0(W_1 \cap W_2^\prime) \subset \Fs_0(W_1 \cap W_2^\prime)$
in this case. The situation is different, however, in $d=2$
spacetime dimensions. 

\begin{prop} \label{prop6.5}
Let $d = 2$, let $\Ws_0$ be the set of all wedges in 
$\RR^2$ which are coherent with a given wedge $W_0$ 
and let $\overline{W_2} \subset W_1 \in \Ws_0$.
Then \\[1.5mm]
(a) \ $\overline{\As_0(W_1 \cap W_2^\prime) \, \Omega} = \Hs$, \\[1.5mm]
(b) \ $ \bigvee_{\, W_1 \cap W_2^\prime \, \subset \, W_0} \, \As_0(W_1 \cap
W_2^\prime) = \Rs(W_0)$, \\[1mm]
(c)  \ $ \bigvee_{\,  W_1 \cap W_2^\prime \, \subset \, W_0^\prime}
\, \As_0(W_1 \cap W_2^\prime) = \widehat{\Rs}(W_0^\prime)$. \\[1.5mm]
Thus one recovers the 
original states and wedge algebras from the net $\As_0$.
\end{prop} 
\begin{proof}
The crucial step in the argument is the demonstration that
the vacuum vector $\Omega$ is cyclic for the algebras 
$\As_0(W_1 \cap W_2^\prime)$. Instead of proving
the cyclicity of $\Omega$ by abstract arguments as in \cite{Le3},
we explicitly exhibit sufficiently many 
operators in $\As_0(W_1 \cap W_2^\prime)$. 
The essential ingredient is the observation  \cite{Le2} that the algebras 
$\Rs(W_1), \Rs(W_2)$ form 
a ``split inclusion'' in $d=2$ dimensions, \ie there
is a von Neumann algebra $\Ms \subset \Bs(\Hs)$ of type I$_\infty$
such that $\Rs(W_2) \subset \Ms \subset \Rs(W_1)$. It then follows
from results of Doplicher and Longo \cite{DL}, cf.\ also 
\cite{BDL}, that there exists a self-adjoint idempotent operator 
$Z_1 \in \Rs(W_1)$ which implements the adjoint action of
$Z$ on $\Rs(W_2)$ and satisfies $Z_1 Z = Z Z_1$,
\ie it is even. Thus if $h$ is any function as
in Lemma \ref{lemma6.2}, one has $ Z_1 \phi(h) \in \Rs(W_1)$. 
On the other hand, if $\text{supp} f \subset W_2$ then
$ Z_1 \phi(h) \phi(f) = -  Z_1 \phi(f) \phi(h)
=  \phi(f)  Z_1 \phi(h)$ and consequently 
$ Z_1 \phi(h) \in \Rs(W_2)^\prime$.

Now let $\Phi \in \Hs$ be orthogonal to $\As_0(W_1 \cap W_2^\prime) \, 
\Omega$. If $h_1, \dots h_{2n + 1}$ are functions as 
in Lemma \ref{lemma6.2}, then 
$Z_1 \phi(h_1) \cdots \phi(h_{2n + 1}) \in \As_0(W_1 \cap W_2^\prime)$
by the preceding argument and consequently $Z_1 \Phi$
must be an element of the even subspace $\Hs_e$. Similarly, if
$h_1, \dots h_{2n}$ are functions as in Lemma \ref{lemma6.2}, 
one has $\phi(h_1) \cdots \phi(h_{2n}) \in \As(W_1 \cap W_2^\prime)$
and consequently $\Phi$ must lie in the subspace  $\Hs_o$
generated by states with an odd particle number. As $Z_1$
is even, this is only possible if $\Phi = 0$.
This establishes the first part of the statement. 
The remaining statements then
follow as in the preceding proposition. 
\end{proof}

In contrast to the situation in higher dimensions, the local and covariant 
net $\As_0$ has the vacuum $\Omega$ as a cyclic vector if $d=2$.
In particular, there exist local operators in  $\As_0$ interpolating
between $\Omega$ and the single particle states. 
One can therefore apply Haag--Ruelle collision theory and finds \cite{Le3}
that the net describes a Boson with nontrivial
scattering matrix $S = (-1)^{N(N-1)/2}$. On the other hand,
the field net $\Fs_0$ defined in relation (\ref{6.4}) coincides with
the net generated by a local free Fermi field and therefore 
describes a Fermion with trivial scattering matrix. Thus, 
in $d=2$ dimensions, the nonlocal net $\Rs$ comprises
different local structures.

\section{Final Comments}

     In a number of recent papers
\cite{Sch1,Sch2,AdS,LoRe,BuLe,Le,Le2,Le3}, it has proven advantageous
to consider nonlocal but weakly local fields for the construction of
local observable algebras by purely algebraic means. Up to now, these
investigations focussed on models in $d = 2$ dimensions. We have
therefore studied in the present paper a prototype of such models in
an arbitrary number of spacetime dimensions. As we have seen, this
model has many features in common with local theories. In fact, there
exist nets $\As_0$, $\Fs_0$ of operators embedded in the original
nonlocal net $\Rs$ which are localized in point--, string-- or
brane--shaped subregions of spacetime, respectively, and (anti)commute
at spatial distances.
 
Within the general setting of algebraic quantum field theory, such
partial locality properties of operators interpolating between the
vacuum and single particle states are sufficient in order to establish
the existence of collision states and of a corresponding scattering
matrix. Moreover, this scattering matrix has the macroscopic causality
(clustering) properties familiar from local field theory.  We refrain
from proving here these statements and refer the interested reader to
\cite{BoBuSch}, where the essential ingredients for a collision theory
involving infinitely extended local operators can be found.

     We have also seen that the net $\Rs$ is maximally nonlocal in
terms of the quantitative measure $\Cs$ introduced in
\cite{BSloc}. Nonetheless, it contains the above well-behaved local
structures and manifests a number of physically desirable
properties. At this stage of our investigation into
locality/nonlocality in quantum field theory, this seems to be a
striking fact.

In view of these findings and the growing interest in nonlocal quantum
field theoretical models, it seems worthwhile to study more
systematically the structure of such theories.  As we have seen in
Section \ref{sec6}, an essential step in the analysis is the
determination of the relative commutants of given inclusions of
algebras.  In general, there is no reason for such relative commutants
to be nontrivial, much less to be large. In the model at hand we have
been able to determine the size of these commutants by explicit
computations. Yet, in a more general setting, it would be desirable to
establish criteria which guarantee their nontriviality. The results in
\cite{AdS,LoRe,BrGuLo,BuLe,Le,Le2,Le3,MSY}, which were also partly
used in the present investigation, indicate that relevant information
to that effect is contained in the modular structure of the
inclusions, although the criterion of modular nuclearity, 
put forward in \cite{BuLe}, seems to be too stringent. It would be
desirable to replace it by a less restrictive condition of a similar
nature which covers a wider range of examples and can be checked more
easily in models.

\newpage

\noindent {\Large \bf Acknowledgements} \\[2mm]
DB wishes to thank the Institute for Fundamental Theory and the
Department of Mathematics of the University
of Florida, and SJS wishes to thank the Institute
for Theoretical Physics of the University of G\"ottingen 
for hospitality and financial support which facilitated this research.
This work was supported in part by a research grant of 
Deutsche Forschungsgemeinschaft (DFG).

\end{document}